\documentclass{PoS}
 \usepackage{amsmath}
\usepackage{subfig}
\usepackage{textcomp}
\usepackage{graphicx}
\usepackage{subfig}
\title{New Noise Subtraction Methods in Lattice QCD}

\ShortTitle{New Noise Subtraction Methods in Lattice QCD}

\author{\speaker{Suman  Baral}\\
        Department of Physics, Baylor University, Waco, TX 76798-7316,USA\\
        E-mail: \email{suman\_baral@baylor.edu}}
\author{Walter Wilcox \\
Department of Physics, Baylor University, Waco, TX 76798-7316, USA \\
E-mail: \email{walter\_wilcox@baylor.edu}}
\author{Ronald B. Morgan \\
Department of Mathematics, Baylor University, Waco, TX 76798-7328, USA \\
E-mail: \email{ronald\_morgan@baylor.edu}}

\abstract{Noise subtraction techniques can help reduce the statistical uncertainty in the extraction of hard to detect signals. We describe new noise subtraction methods in Lattice QCD which apply to disconnected diagram evaluations. Some of the noise suppression techniques include polynomial quark matrix methods, eigenspectrum deflation methods, and combination methods. Our most promising technique combines polynomial and Hermitian deflation subtraction methods. The overall goal is to improve the efficiency of Lattice QCD noise method algorithms.}

\FullConference{34th annual International Symposium on Lattice Field Theory\\
		24-30 July 2016\\
		University of Southampton, UK}

\begin{document}

\section{Introduction}
Lattice QCD is a set of numerical techniques which uses a finite space-time lattice to simulate the interactions between quarks and gluons. The evaluation of quark loop effects on a given lattice is essential but extremely computer time intensive and approximation techniques must be introduced\cite{Wilcox}. In this paper, we describe some new noise subtraction methods useful in evaluating quark operators. Perturbative subtraction\cite{Thron} is a standard method which we will be comparing to and attempting to improve upon. This paper will be focusing on eigenspectrum subtraction (deflation), polynomial subtraction, and combination methods.


\section{Methods}
Many operators in lattice QCD simulations are flooded with noise. Several strategies could be applied to reduce the variance of these operators, which originates in the off-diagonal components of the associated quark matrix. One basic strategy is to mimic the off-diagonal elements of the inverse of the quark matrix with another traceless matrix, thereby maintaining the trace but with reduced statistical uncertainty. We have applied several new techniques to non-subtracted (NS) lattice data. These are termed eigenvalue subtraction (ES)\cite{Guerrero}, Hermitian forced eigenvalue subtraction (HFES)\cite{Guerrero2} and polynomial subtraction (POLY)\cite{Liu}. In \cite{Guerrero2} we also introduced techniques which combine deflation with other subtraction methods. Here we will compare the various methods and show how effective the combination methods are. We work with the standard Wilson matrix in the quenched approximation. The size of the lattice we used is $24^{3}\times 32$, the number of noises is 200, the kappa value is 0.155, and we use $Z4$ noise. Linear equations are solved using GMRES-DR (generalized minimum residual algorithm-deflated and restarted) for the first noise and GMRES-Proj (similar algorithm projected over eigenvectors) for remaining noises\cite{Morgan} .

In order to calculate the trace, the quark matrix $M$ is projected over a finite number of noises $\eta^{(n)}$, $Mx^{(n)}=\eta^{(n)}$, and solution vectors $x^{(n)}$ are extracted. All of our methods attempt to design a traceless matrix $\tilde M^{-1}$ in order to obtain off-diagonal elements as close to  $M^{-1}$ as possible. We then use solution vectors formed from such a matrix. Unfortunately, this matrix is not completely traceless, so we will have to re-add the appropriate trace. For the trace of the inverse quark matrix one has
 \begin{equation}
Tr\left(M^{-1}\right)=\frac{1}{N}\sum_n^N{\left( \eta ^{(n)\dagger} \left(  x^{(n)} -\tilde x^{(n)}\right) \right)} + Tr\left(  \tilde M^{-1}\right),
\end{equation}
for $N$ noises, where $x^{(n)}$ is the solution vector generated when implementing the GMRES algorithms and $\tilde x^{(n)}$ is given by
\begin{equation}
\tilde x^{(n)}\equiv \tilde M^{-1}\eta^{(n)}.
\end{equation}
For any operator $\Theta$, the appropriate trace becomes
\begin{equation}
Tr\left( \Theta M^{-1}\right)=\frac{1}{N}\sum_n^N{\left( \eta ^{(n)\dagger} \Theta \left(x^{(n)} -\tilde x^{(n)}\right) \right)} + Tr\left( \Theta \tilde M^{-1}\right).
\end{equation}
Note that adding $Tr\left(\Theta  \tilde M^{-1}\right)$ has no influence on the noise error bar, which is what is studied here.

\subsection{Eigenvalue Subtraction (ES)}
The spectrum of low eigenvalues of matrices can limit the performance of iterative solvers. We have emphasized the role of deflation in accelerating the convergence of algorithms in Ref.\cite{Wilcox2}. Here we investigate deflation effects in statistical error reduction. Consider the vectors $e_R^{(q)}$ and $e_L^{(q)\dagger}$, which are defined as normalized right and left eigenvectors of the matrix $M$, as in
\begin{equation}
Me_R^{(q)}=\lambda^{(q)} e_R^{(q)},
\end{equation}
and
\begin{equation}
 e_L^{(q)\dagger} M=\lambda^{(q)} e_L^{(q)\dagger},
\end{equation}
where $\lambda^{(q)}$ is the eigenvalue associated with both eigenvectors. With a full set $N$ of eigenvectors and eigenvalues the matrix $M$ can be fully formed as 
\begin{equation}
M=\sum_{q=1}^{N} {e_R^{(q)}  \lambda^{(q)} e_{L}^{(q)\dagger}},
\end{equation}
or
\begin{equation}
M=V_R \Lambda V_L^{\dagger}, 
\end{equation}
where $V_R$ contains the right eigenvectors and $V_L^{\dagger}$ contains the left eigenvectors. $\Lambda$ is a purely diagonal matrix made up of the eigenvalues of $M$ in the order they appear in both $V_R$ and $V_L$. Deflating out eigenvalues with the linear equation solver GMRES-DR can mimic the low eigenvalue structure of the inverse of matrix $M$ as
\begin{equation} 
\tilde{M}_{eig}^{-1} \equiv \tilde{V}_{R} \tilde{\Lambda}^{-1} \tilde{V}_{L}^{\dagger},
\end{equation}
where $\tilde{V}_{R}$ and  $\tilde{V}_{L}^{\dagger}$  are the computed right and left eigenvectors and $ \tilde{\Lambda}^{-1}$ is the inverse of eigenvalues. 

In the results we will see that if we n\"aively subtract eigenvalues from a non-Hermitian matrix, we often end up expanding the size of error bars. This happens because many of the right handed eigenvectors of a non-Hermitian matrix can point in the same direction, a condition referred to as \lq\lq highly non-normal".

The trace takes the following form in this method,
\begin {equation}
Tr\left( \Theta M^{-1}\right)=\frac{1}{N}\sum_n^N{\left( \eta ^{(n)\dagger}  \Theta \left(  x^{(n)} -\tilde x^{(n)}_{eig}\right) \right)} + Tr\left( \Theta \tilde M^{-1}_{eig}\right),
\end {equation}
where $\tilde x^{(n)}_{eig} = \tilde M_{eig}^{-1}\eta^{(n)}$. This last operation does not add matrix vector products. The generation of eigenmodes only requires the super convergence solution of a single right hand side with GMRES-DR. As pointed out previously\cite{Guerrero}, there is a relation between the even-odd eigenvectors for the reduced system and the full eigenvectors. Other right hand sides are accelerated with GMRES-Proj using the eigenvalues generated.


\subsection{Hermitian Forced Eigenvalue Subtraction (HFES)}
To avoid the non-normal problem we force our matrix to be formulated in a Hermitian manner. The easiest way for us to do this with the Wilson matrix is to multiply by the Dirac $\gamma_5$ matrix. It is important for the algorithm to do the multiplication on the {\it right}, $M\gamma_5$, to avoid using cyclic properties which fail in finite noise space. We can then form the low eigenvalue structure of $M\gamma_5$ from these eigenvalues. We define 
\begin{equation}
M'\equiv M\gamma_5.
\end{equation}
We can now form normalized eigenvectors ${e'}_R^{(n)}$, eigenvalues $\lambda '^{(n)}$ and solution vectors $\tilde{x}'{_{eig}^{(n)}}$ for this new Hermitian matrix and perform a calculation similar to the ES method, accounting properly for the extra $\gamma_5$ factors. The trace of any operator  $\Theta$, for the HFES method takes the following form,
\begin {equation}
Tr\left( \Theta M^{-1}\right)=\frac{1}{N}\sum_n^N{\left( \eta ^{(n)\dagger}  \Theta \left(  x^{(n)} -\tilde{x}'{_{eig}^{(n)}}\right) \right)} + Tr\left( \Theta \gamma_5\tilde {M}'{_{eig}^{-1}}\right),
\end {equation}
where
\begin{equation}
\tilde{x}'{_{eig}^{(n)}} \equiv\gamma_5\tilde{M}'{_{eig}^{-1}}\eta^{(n)}
                           =\gamma_5 \sum_{q}^{Q}{\frac{1}{{\lambda ' }^{(q)}}  { e'}_{R}^{(q)}  \left({e'}_{R}^{(q)\dagger} \eta^{(n)} \right) }
\end{equation}
and 
\begin{equation}
\tilde{M}'{_{eig}^{-1}} \equiv \tilde{V}_{R}' \tilde{\Lambda'}^{-1} \tilde{V}_{R}'^{\dagger}.
\end{equation}
$\tilde{V}_{R}'$ is a matrix whose columns are the Q smallest right eigenvectors of $M'$. $\tilde{\Lambda '}^{-1}$ is the diagonal matrix of size Q that contains the inverse of eigenvalues $1/{{\lambda'}^{(q)}}$ as the diagonal elements. The price paid here, similar to the ES method, is a single extra super convergence on one right hand side for the non-reduced Hermitian system $M'$ with GMRES-DR to extract eigenvectors and eigenvalues.


 \subsection{Polynomial Subtraction (POLY)} 
Our goal is to find more efficient methods than perturbative subtraction (PS), where to 6th order:
\begin{equation}
\tilde{M}_{pert}^{-1} \equiv 1+\kappa P+(\kappa P)^2+(\kappa P)^3+(\kappa P)^4+(\kappa P)^5+(\kappa P)^6.
\end{equation}
$P$ is the quark hopping matrix and $\kappa$ is the usual expansion parameter. The polynomial method is similar to that of perturbative subtraction. The only difference is that the coefficients are allowed to be different from one,
\begin{equation}
\tilde{M}_{poly}^{-1} \equiv a_1+ a_2\kappa P+a_3(\kappa P)^2+a_4(\kappa P)^3+a_5(\kappa P)^4+a_6(\kappa P)^5+a_7(\kappa P)^6,
\end{equation}
where the $a_i$'s are the coefficients obtained from min-res projection\cite{Liu}. The trace in this method takes the form,
\begin {equation}
Tr\left( \Theta M^{-1}\right)=\frac{1}{N}\sum_n^N{\left( \eta ^{(n)\dagger} \cdot \Theta \left(  x^{(n)} -\tilde x^{(n)}_{poly}\right) \right)} + Tr\left( \Theta \tilde M^{-1}_{poly}\right),
\end {equation}
where
\begin{equation}
\tilde x^{(n)}_{poly}\equiv\tilde M^{-1}_{poly}\eta^{(n)}.
 \end{equation}
 

\subsection{Combination Methods (HFPOLY and HFPS)}
We have developed two methods which combine the error reduction techniques of HFES with POLY and PS, called HFPOLY and HFPS. N\"aively, for POLY we could think of this method as a subtracted combination: $\tilde M^{-1}_{poly}+\gamma_5 \tilde M'^{-1}_{eig}$. However, this presents a possible conflict since $\tilde M^{-1}_{poly}$ will overlap on the deflated Hermitian eigenvector space. In order to prevent this, we also remove low eigenmode information from $\tilde {M}^{-1}_{poly}$. Since $\tilde {M}_{poly}$ is not Hermitian, the procedure is to define $\tilde M'_{poly}=\tilde {M}_{poly}\gamma_5$ and remove its overlapping Hermitian eigenvalue information using the eigenvectors from $M'$. Following the idea in Ref.\cite{Guerrero}, we define
\begin{equation}
{e'}_R^{(q)\dagger}\tilde M'^{-1}_{poly}{e'}_R^{(q)}\equiv\frac{1}{\xi'^{(q)}},
\end{equation}
where ${e'}^{(q)}_{R}$ is the eigenmode of $M'$ generated within HFES method and $1/\xi'^{(q)}$ are the approximate eigenvalues of $\tilde M'^{-1}_{poly}$. The trace takes the following form,
 \begin{eqnarray}                                 
Tr\left(  \Theta M^{-1}\right ) = \frac{1}{N} & \sum_n^N { \left( \eta^{(n)\dagger} \left[  \Theta x^{{(n)}}  - \Theta\tilde{x}'{_{eig}^{(n)}} - \left( \Theta \tilde{x}_{poly}^{(n)}  - \Theta \tilde{x}'{_{eigpoly}^{(n)}} \right)  \right ] \right) } \nonumber\\+& Tr \left(  \Theta \gamma_5 \tilde{M'}_{eig}^{-1}\right)+Tr\left( \Theta \tilde{M}_{poly}^{-1}- \Theta \gamma_5 \tilde{M'}_{eigpoly}^{{-1}} \right) ,
\end{eqnarray}
where
 \begin{equation}
        \tilde{x}'{_{eigpoly}^{(n)}} \equiv\gamma_5\tilde{M}'{_{eigpoly}^{-1}}\eta^{(n)}
                           =\gamma_5 \sum_{q}^{Q}{\frac{1}{\xi'^{(q)}}  { e'}_{R}^{(q)}  \left({e'}_{R}^{(q)\dagger} \eta^{(n)} \right) }.
\end{equation}
$\tilde{x}'{_{eig}^{(n)}}$ and $\tilde x^{(n)}_{poly}$ are defined in previous sections and
\begin{equation}
\tilde{M}_{eigpoly}^{'^{-1}}  \equiv \tilde{V}_{R}' \Xi^{-1} \tilde{V}_{R}'^{\dagger},
\end{equation}
where $\tilde{V}_{R}'$ is defined above also and $\Xi^{-1}$ is the diagonal matrix of size Q that contains approximate inverse eigenvalues, $1/\xi'^{(q)}$. 

In the case of HFPS, $\tilde{M}_{poly}^{-1}$ is replaced by $\tilde{M}_{pert}^{-1}$ and all the calculations are repeated.
     

\section{ Results}
Figure 1 shows the calculated error bars for the nonlocal current operator in the one direction (other currents are similar) as a function of deflated eigenvectors. Similarly, Figure 2 shows error bars for the local current operator in the one direction. Figure 3 represents the error bar for the scalar operator. We deflated approximately 140 eigenvectors.

The ES method does not decrease the error bars as the number of deflated eigenvectors is increased. This arises from the non-normal nature of the non-Hermitian eigenvectors\footnote{Note that the numerical ES results in Ref.\cite{Guerrero} were in error; see Ref.\cite{Guerrero2}}. The HFES method reduces the error bars in all cases. However it does not outperform PS for the number of eigenvectors removed. The POLY method is better than PS for nonlocal and scalar operators but the difference is not that significant for local operators. The HFPOLY combo is the most efficient method. We define the relative efficiency, $RE$, of the two methods as
\begin{equation}
RE\equiv \left( \frac {1}{\delta y^{2}}-1\right) \times 100,
\end{equation}
where $\delta y$ is the relative error bar. The relative error bars for HFPOLY combo as compared to PS are approximately $0.77, 0.75$ and $0.74$ for scalar, local and nonlocal operators, respectively. That means the HFPOLY method is more efficient than the PS method by 68\%, 77\% and 81\%, respectively. We expect efficiency to improve further as we move on towards lower quark masses. Note that similar results for deflation applied to hierarchical probing have been obtained in Ref.\cite{Gambhir}.
 
\begin{figure}[!htpb]
\centering
\includegraphics[width=.75\textwidth]{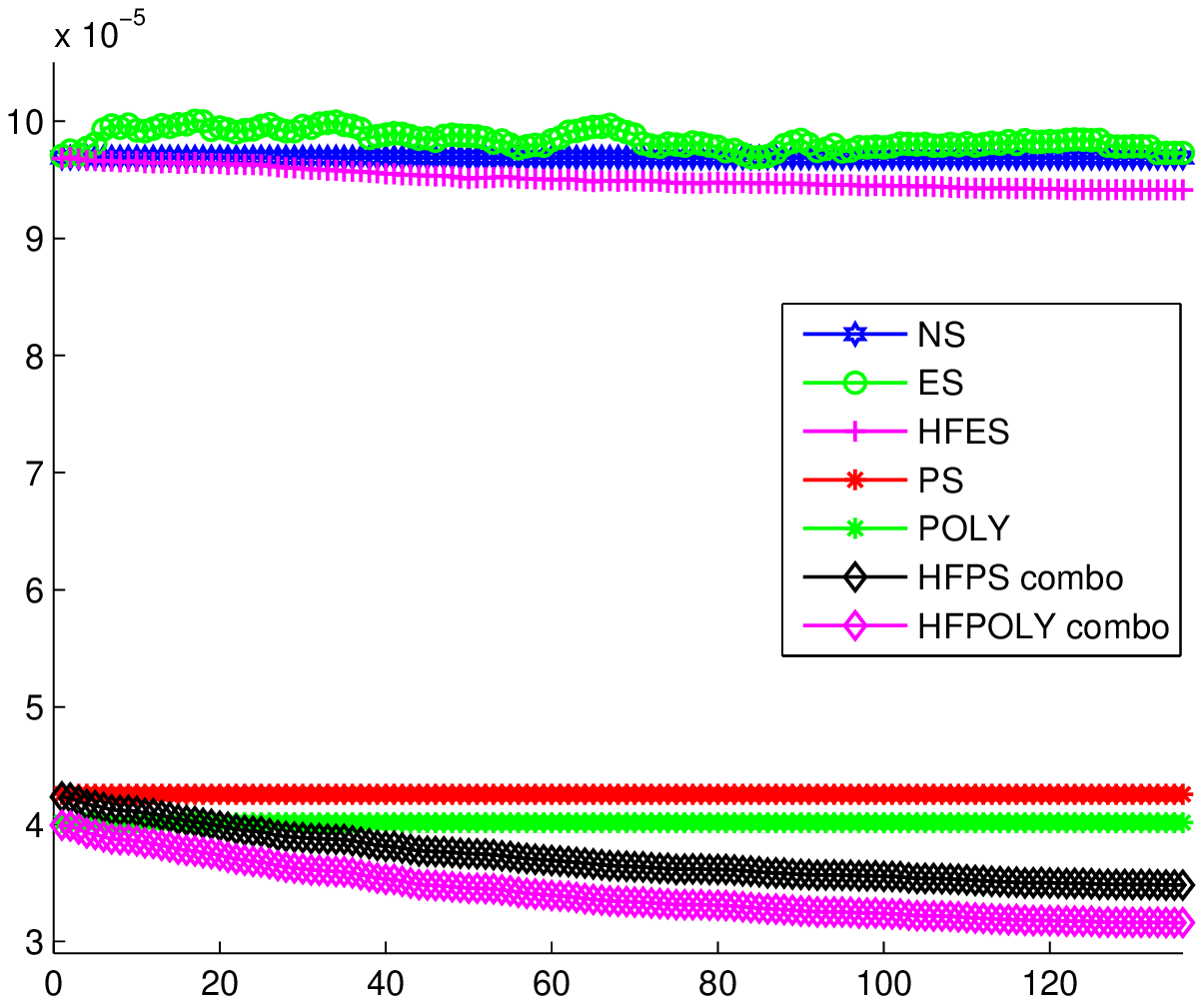}
    
\caption{Error bars for a nonlocal spatial vector as a function of deflated eigenvalues.}\label{fig:1}

\includegraphics[width=.75\textwidth]{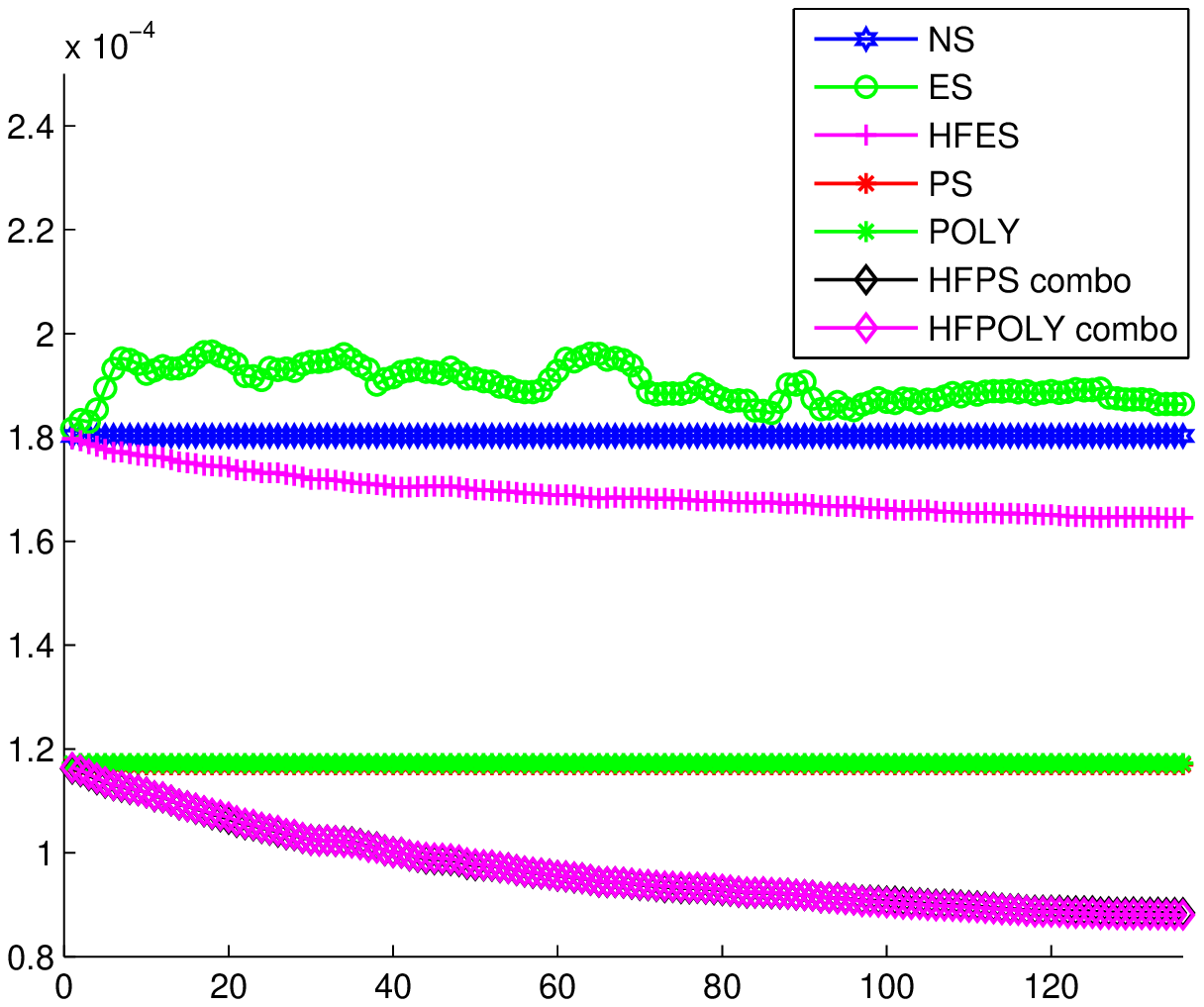}

\caption{Error bars for a local spatial vector as a function of deflated eigenvalues.}\label{fig:2}
\end{figure}

\begin{figure}[!htpb]
\centering
\includegraphics[width=.75\textwidth]{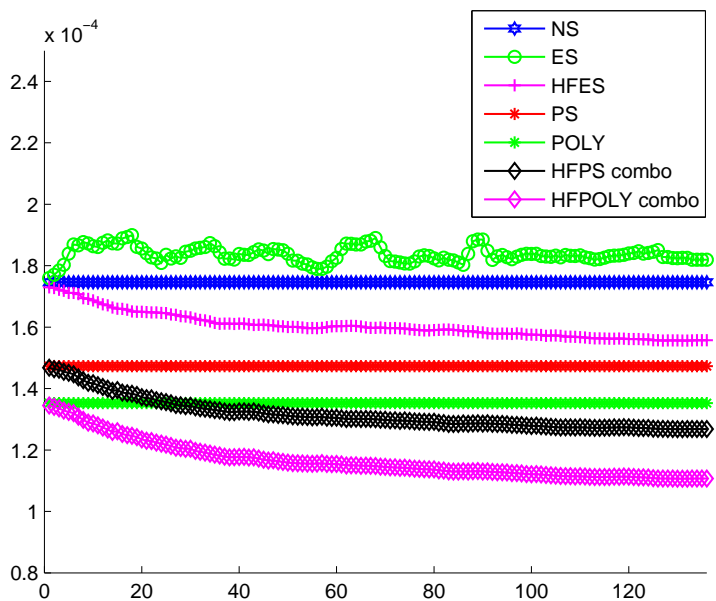}
\caption{Error bars for the scalar operator as a function of deflated eigenvalues.}
\label{fig3}
\end{figure}
 

\section{Conclusions}
Our polynomial and perturbative deflation combination methods produce very encouraging results for $\kappa=0.155$. Although our quark mass is not small in this investigation, we are hopeful that our methods will be effective at smaller quark mass.  
 

\section{Acknowledgements} 
All of the numerical work was performed using the High Performance Cluster at Baylor University. This work is partially supported by URC grant funding, Baylor University.

\bibliographystyle{plain}
\bibliography{sample}
\end{document}